\documentstyle[12pt]{article}
\newcommand{\al}{\alpha}
\newcommand{\ba}{\begin{array}}
\newcommand{\be}{\begin{equation}}
\newcommand{\bea}{\begin{eqnarray}}
\newcommand{\cc}{{\cal C}}
\newcommand{\ck}{{\cal K}}
\newcommand{\cl}{{\cal L}} 
\newcommand{\da}{\dagger}
\newcommand{\ddz}{\frac{d}{dz}}
\newcommand{\ddzt}{\frac{d^2}{dz^2}}
\newcommand{\ddzth}{\frac{d^3}{dz^3}}  
\newcommand{\ea}{\end{array}}
\newcommand{\ee}{\end{equation}}
\newcommand{\eea}{\end{eqnarray}}
\newcommand{\fr}{\frac}

\newcommand{\lb}{\label}
\newcommand{\ld}{\ldots}
\newcommand{\lcb}{\left\{}
\newcommand{\lla}{\left\langle}
\newcommand{\lmd}{\left|}
\newcommand{\lra}{\longrightarrow} 
\newcommand{\lrb}{\left(}
\newcommand{\lsb}{\left[}
\newcommand{\nn}{\nonumber}

\newcommand{\rmd}{\right|}
\newcommand{\rra}{\right\rangle}
\newcommand{\rcb}{\right\}}
\newcommand{\rrb}{\right)}
\newcommand{\rsb}{\right]}

\begin{document}

\baselineskip16pt

\begin{center}
{\LARGE \bf Quadratic algebras\,:Three-mode bosonic realizations and
applications} \\
\end{center}

\smallskip

\baselineskip14pt

\begin{center}
V. Sunil Kumar$^1$, B. A. Bambah$^2$, and R. Jagannathan$^3$ \\
{\em $^{1,2}$School of Physics, University of Hyderabad \\ 
Hyderabad - 500046, India \\
$^3$The Institute of Mathematical Sciences \\ 
C.I.T. Campus Tharamani, Chennai - 600113, India}  
\end{center}

\vspace{0.5cm}

\noindent
Quadratic algebras of the type $\lsb Q_0 , Q_\pm \rsb$ $=$ $\pm Q_\pm$, 
$\lsb Q_+ , Q_- \rsb$ $=$ $aQ_0^2 + bQ_0 + c$ are studied using three-mode 
bosonic realizations.  Matrix representations and single variable differential 
operator realizations are obtained.  Examples of physical relevance of such 
algebras are given. 

\vspace{1cm}

\section{Introduction}
\renewcommand{\theequation}{\arabic{section}.{\arabic{equation}}}
\setcounter{equation}{0}
In recent times there has been a great deal of interest in non-linear  
deformations of Lie algebras because of their significant applications 
in several branches of physics.  This is largely based on the realization 
that the physical operators relevant for defining the dynamical algebra 
of a system need not be closed under a linear (Lie) algebra, but might 
obey a nonlinear algebra.  Such nonlinear algebras are, in general, 
characterized by commutation relations of the form 
\be
\lsb T_i , T_j \rsb = C_{ij} \lrb T_k \rrb\,, 
\lb{cr} 
\ee 
where the functions $C_{ij}$ of the generators $\lcb T_k \rcb$ are constrained 
by the Jacobi identity 
\be
\lsb T_i , C_{jk} \rsb + 
\lsb T_j , C_{ki} \rsb + 
\lsb T_k , C_{ij} \rsb = 0\,.  
\ee
The functions $C_{ij}$ can be infinite power series in $\lcb T_k \rcb$ as is 
in the case of quantum algebras (with further Hopf algebraic restrictions) 
and $q$-oscillator algebras.  When $\lcb C_{ij} \rcb$ are polynomials of the 
generators one gets the socalled polynomially nonlinear, or simply polynomial, 
algebras.  A special case of interest is when the commutation relations 
(\ref{cr}) take the form 
\be
\lsb T_i , T_j \rsb = c_{ij}^k T_k\,, \quad 
\lsb T_i , T_\al \rsb = t_{i\al}^\beta T_\beta\,, \quad 
\lsb T_\al , T_\beta \rsb = f_{\al\beta} \lrb T_k \rrb\,,  
\ee 
containing a linear subalgebra.  Simplest examples of such algebras occur 
when one gets 
\be
\lsb N_0 , N_\pm \rsb = \pm N_\pm\,, \qquad
\lsb N_+ , N_- \rsb = f\lrb N_0 \rrb\,. 
\lb{fn} 
\ee 
In general, the Casimir operator of this algebra (\ref{fn}) is seen to be 
given by 
\be 
\cc = N_+ N_- + g \lrb N_0 - 1 \rrb = N_- N_+ + g \lrb N_0 \rrb\,, 
\lb{genC}
\ee
where $g \lrb N_0 \rrb$ can be determined from the relation 
\be
g \lrb N_0 \rrb - g \lrb N_0 - 1 \rrb = f \lrb N_0 \rrb\,.
\lb{ggf}
\ee  

If $f\lrb N_0 \rrb$ is quadratic in $N_0$ we have a quadratic algebra and  
if $f\lrb N_0 \rrb$ is cubic in $N_0$ we have a cubic algebra.  The nonlinear 
algebras, in particular the quadratic and cubic algebras, and their 
representations have been studied in connection with several problems in 
quantum mechanics, statistical physics, field theory, Yang-Mills type 
gauge theories, two-dimensional integrable systems, etc. (\cite{L}-\cite{F}).  
As in the case of the Lie algebras, compact nonlinear algebras have finite 
dimensional unitary irreducible representations and noncompact algebras have 
infinite dimensional unitary irreducible representations.  These nonlinear 
algebras have also been found to be very useful in quantum optics with the 
observation that quantum optical Hamiltonians describing multiphoton processes 
have symmetries described by polynomially deformed $su(2)$ and $su(1,1)$ 
algebras \cite{K}.  Coherent states of different kinds of nonlinear oscillator 
algebras have been presented by several authors (\cite{M}-\cite{Q2}).  
Recently, a general unified approach for finding the coherent states of 
polynomial algebras, relevant for various multiphoton processes in quantum 
optics, has been presented by us \cite{Su}.  \\ 

Algebras of the type (\ref{fn}) have been studied in mathematics literature 
also and shown to have a surprisingly rich theory of representations \cite{Sm}. 
Algebras of the type (\ref{fn}) with commutator $\lsb N_+ , N_- \rsb$ replaced 
by the anticommutator $\lcb N_+ , N_- \rcb$ leading to polynomial deformations 
of the superalgebra $osp(1|2)$ have also been investigated \cite{Jv}.  As is 
well known, in the case of classical Lie algebras bosonic realizations play a 
very useful role in the representation theory and applications to physical 
problems. The main purpose of this work is to study some aspects of quadratic 
algebras relating to three-mode bosonic realizations, and the associated matrix 
representations, differential operator realizations, coherent states, and 
physical applications.  In Section 1 we briefly review the two-mode bosonic 
construction of $su(2)$ and $su(1,1)$ algebras.  Following closely our earlier 
work \cite{Su}, in Sections 2 and 3 we construct the compact and noncompact 
quadratic algebras, respectively, using three bosonic modes and study the 
associated representations.  In Section 4 we briefly discuss the coherent 
states of these algebras.  Finally, in Section 5 we conclude with a few remarks 
on the physical and mathematical significance of these quadratic and higher 
order nonlinear algebras.   

\section{Two-mode construction of $su(2)$ and $su(1,1)$\,: A brief review} 
\renewcommand{\theequation}{\arabic{section}.{\arabic{equation}}}
\setcounter{equation}{0}

Let us briefly recall the study of $su(2)$ and $su(1,1)$ in terms of two-mode 
bosonic realizations, to fix the framework and notations for our work.  The 
$su(2)$ algebra, 
\be
\lsb J_0 , J_\pm \rsb = \pm J_\pm\,, \quad
\lsb J_+ , J_- \rsb = 2J_0\,, 
\ee
is satisfied by the Jordan-Schwinger realization 
\be
J_0 = \fr{1}{2} \lrb a_1^\da a_1 - a_2^\da a_2 \rrb\,, \quad 
J_+ = a_1^\da a_2\,, \quad
J_- = J_+^\da = a_1 a_2^\da\,.  
\lb{js}
\ee 
In this realization the Casimir operator becomes  
\be
\cc = J^2 = J_+J_- + J_0 \lrb J_0 - 1 \rrb 
    = \fr{1}{4} \lrb a_1^\da a_1 + a_2^\da a_2 \rrb 
                \lrb a_1^\da a_1 + a_2^\da a_2 + 2 \rrb\,.
\ee
Consequently, the application of the realization (\ref{js}) on a set of $2j+1$   
two-mode Fock states $\lmd n_1,n_2 \rra$ with constant $n_1+n_2 = 2j$ leads to 
the $(2j+1)$-dimensional unitary irreducible representation for each  
$j = 0,1/2,1,\ld\,$. Thus, with 
$\lcb \lmd j,m \rra = \lmd n_1 = j+m, n_2=j-m \rra \lmd \right. m = j,j-1,
\ld\,,-j \rcb$ as the basis states, one gets the $j$-th unitary irreducible  
representation 
\bea
J_0 \lmd j , m \rra & = & m \lmd j , m \rra\,, \nn \\ 
J_\pm \lmd j , m \rra & = & 
      \sqrt{(j \mp m)(j \pm m+1)}\,\lmd j , m \pm 1 \rra\,, \nn \\ 
   &  & \qquad \qquad m = j,j-1,\ld\,,-j\,, 
\lb{mrj}
\eea 
associated with the constant value $j(j+1)$ for the Casimir operator $J^2$.  \\  

Let us now consider the single variable differential operator realization 
corresponding to the above matrix representation (\ref{mrj}).  With the 
Fock-Bargmann correspondence 
\be
\lrb a^\da , a \rrb \lra \lrb z , \ddz \rrb\,, \qquad
\lmd n \rra \lra \fr{z^n}{\sqrt{n!}}\,,
\ee
we can make the association 
\bea
\lmd j , m \rra & \lra & \fr{z_1^{j+m} z_2^{j-m}}{\sqrt{(j+m)!(j-m)!}} 
  = \fr{z_2^{2j} \lrb z_1/z_2 \rrb^{j+m}}{\sqrt{(j+m)!(j-m)!}}\,, \nn \\
  &  &  \qquad \qquad m = -j, -j+1, \ld\,, j-1, j\,.
\eea  
Since $j$ is a constant for a given representation we can rewrite the above 
as a mapping to monomials 
\be 
\lmd j , m \rra \lra \psi_{j,n}(z) = \fr{z^n}{\sqrt{n!(2j-n)!}}\,, \quad 
n = 0, 1, 2, \ld\,, 2j\,. 
\lb{mono} 
\ee 
Then, it is obvious that the above set of $(2j+1)$ monomials (\ref{mono}) 
forms the basis carrying the finite dimensional representation (\ref{mrj}) 
corresponding to the single variable realization 
\be 
J_0 = z \ddz - j\,, \quad 
J_+ = -z^2 \ddz + 2jz\,, \quad 
J_- = \ddz\,. 
\ee

In an analogous way, for $su(1,1)$ the two-mode bosonic realization 
\be
K_0 = \fr{1}{2} \lrb a_1^\da a_1 + a_2^\da a_2 + 1 \rrb\,, \quad
K_+ = a_1^\da a_2^\da\,, \quad
K_- = K_+^\da = a_1 a_2\,, 
\lb{su11}
\ee
satisfies the algebra 
\be
\lsb K_0 , K_\pm \rsb = \pm K_\pm\,, \qquad 
\lsb K_+ , K_- \rsb = -2K_0\,.
\ee
The Casimir operator is 
\be 
\cc = K^2 = K_+ K_- - K_0 \lrb K_0-1 \rrb  
    = \fr{1}{4} \lsb 1 - \lrb a_1^\da a_1 - a_2^\da a_2 \rrb^2 \rsb\,.
\ee 
Consequently, the application of the realization (\ref{su11}) on any infinite 
set of two-mode Fock states 
$\lcb \lmd \left. k , n \rra = \lmd n , n+2k-1 \rra 
\rmd n = 0,1,2,\,\ld \rcb$ (or $\lcb \lmd \left. k , n \rra 
= \lmd n+2k-1 , n \rra \rmd n = 0,1,2,\,\ld \rcb$), with constant 
$\lmd n_1 - n_2 \rmd = 2k-1$, leads to the infinite dimensional unitary 
irreducible representation, the so-called positive discrete representation 
${\cal D}^+(k)$\,, for each $k = 1/2, 1, 3/2,\,\ld\,.$\,: 
\bea
K_0 \lmd k , n \rra & = & (k+n) \lmd k , n \rra\,, \nn \\
K_+ \lmd k , n \rra & = & \sqrt{(2k+n)(n+1)}\,\lmd k , n+1 \rra\,, \nn \\
K_- \lmd k , n \rra & = & \sqrt{(2k+n-1)n}\,\lmd k , n-1 \rra\,, \nn \\
  &  & \qquad \qquad n = 0,1,2,\,\ld\,. 
\lb{mrk}
\eea 
with the Casimir operator $K^2$ taking  the value $k(1-k)$ in the  
representation. \\

As in the $su(2)$ case, we can make the association 
\be
\lmd k , n \rra \lra \fr{z_1^n z_2^{n+2k-1}}{\sqrt{n!(n+2k-1)!}} 
     = \fr{\lrb z_1 z_2 \rrb^n z_2^{2k-1}}{\sqrt{n!(n+2k-1)!}}\,.
\ee
Then, with $k$ being constant in a given representation, it is obvious that 
the infinite set of monomials 
\be
\psi_{k,n}(z) = \fr{z^n}{\sqrt{n!(n+2k-1)!}}\,, \quad 
n = 0,1,2,\,\ld\,, 
\ee
forms the basis carrying the representation (\ref{mrk}) corresponding to 
the single variable realization 
\be
K_0 = z \ddz + k\,, \quad 
K_+ = z\,, \quad 
K_- = z \ddzt + 2k \ddz\,.
\ee

\section{Three-mode construction of compact quadratic algebras} 
\renewcommand{\theequation}{\arabic{section}.{\arabic{equation}}}
\setcounter{equation}{0}

\noindent
A quadratic algebra is defined, in general, by the commutation relations 
\be 
\lsb Q_0 , Q_\pm \rsb = \pm Q_\pm\,, \quad 
\lsb Q_+ , Q_- \rsb = a Q_0^2 + b Q_0 + c\,, 
\ee 
where $(a,b,c)$ are constants, or commute among themselves and with 
$\lrb Q_0, Q_+, Q_- \rrb$ so that they take constant values in any irreducible 
representation.  When $a = c = 0$ and $b = \pm 2$ one gets, respectively, 
$su(2)$ and $su(1,1)$ as special cases.  Let us now construct a class of 
compact quadratic algebras using three bosonic modes.  To this end, let 
\bea 
Q_0 & = & \fr{1}{4} \lrb a_1^\da a_1 + a_2^\da a_2 - 2a_3^\da a_3 + 1 \rrb 
 = \fr{1}{2} \lrb K_0 - a_3^\da a_3 \rrb\,, \nn \\ 
Q_+ & = & a_1^\da a_2^\da a_3 = K_+ a_3\,, \qquad 
Q_- = Q_+^\da = a_1 a_2 a_3^\da = K_- a_3^\da\,, \nn \\
\ck & = & \frac{1}{4} 
  \lsb 1 - \lrb a_1^\da a_1 - a_2^\da a_2 \rrb^2 \rsb = K^2\,, \nn \\ 
\cl & = & \fr{1}{4} \lrb a_1^\da a_1 + a_2^\da a_2 + 2a_3^\da a_3 + 1 \rrb 
      = \frac{1}{2} \lrb K_0 + a_3^\da a_3 \rrb\,,    
\lb{l} 
\eea
where $\lrb K_0 , K_+ , K_- \rrb$ generate $su(1,1)$ with $K^2$ as the  
Casimir operator.  Then, it is found that 
\bea
\lsb \ck , \cl \rsb & = & 0\,, \quad 
\lsb \ck , Q_{0,\pm} \rsb = 0\,, \quad 
\lsb \cl , Q_{0,\pm} \rsb = 0\,, \nn \\ 
\lsb Q_0 , Q_\pm \rsb & = & \pm Q_\pm\,, \quad 
\lsb Q_+ , Q_- \rsb = 3Q_0^2 + (2\cl-1) Q_0 
                        + \lrb \ck - \cl (\cl + 1) \rrb\,.\nn \\  
  &   &  
\lb{cqa}
\eea 
Thus, $\lrb Q_0 , Q_+, Q_- \rrb$ generate a quadratic 
algebra with $\ck$ and $\cl$ taking constant values in any irreducible 
representation.  The Casimir operator of this algebra is given by 
\be 
\cc = Q_+ Q_- + Q_0^3 + (\cl - 2) Q_0^2 
        + \lrb \ck - \cl^2 -2\cl + 1 \rrb Q_0\,, 
\ee
apart from an additional constant function of $\ck$ and $\cl$, following 
the recipe contained in (\ref{fn})-(\ref{ggf}). \\ 

The condition that $\ck$ and $\cl$ take constant values in an irreducible 
representation fixes the basis to be the set of three-mode Fock states 
\be
\lmd k , l , n \rra = \lmd n, n+2k-1, 2l-k-n \rra\,, \quad 
 n = 0, 1, 2,\,\ld\,, (2l-k)\,, 
\lb{cqabs} 
\ee
with 
\be
2l-k = 0,1,2,\,\ld\,, \qquad 
k = 1/2, 1, 3/2,\,\ld\,, 
\ee  
and 
\be 
\ck \lmd k , l , n \rra = k(1-k) \lmd k , l , n \rra\,, \quad 
\cl \lmd k , l , n \rra = l \lmd k , l , n \rra\,. 
\ee 
The basis states (\ref{cqabs}) carry the $(2l-k+1)$-dimensional unitary 
irreducible representation of the quadratic algebra (\ref{cqa}) which can 
be labeled by the values of the pair $(k,l)$.  Explicitly, the $(k,l)$-th 
representation is\,:  
\bea 
Q_0 \lmd k , l , n \rra & = & (k-l+n) \lmd k , l , n \rra\,, \nn \\ 
Q_+ \lmd k , l , n \rra & = & 
  \sqrt{(n+1)(n+2k)(2l-n-k)}\,\lmd k , l , n+1 \rra\,, \nn \\
Q_- \lmd k , l , n \rra & = & 
  \sqrt{n(n+2k-1)(2l-n-k+1)}\,\lmd k , l , n-1 \rra\,. \nn \\ 
\ck \lmd k , l , n \rra & = & k(1-k) \lmd k , l , n \rra\,, \quad 
\cl \lmd k , l , n \rra = l \lmd k , l , n \rra\,, \nn \\    
 &  & \qquad \qquad n = 0,1,2,\,\ld\,,(2l-k)\,. 
\lb{mrkl} 
\eea 
The Casimir operator has the value $\lrb l^3 + (l+1)[k(1-k)-1] + 1 \rrb$ in 
this representation.  \\ 

It should be noted that for $k > 1/2$ the same representation is obtained by 
the choice of basis states as 
$\lmd k , l , n \rra = \lmd n+2k-1, n, 2l-k-n \rra\,,$ with 
$n = 0, 1, 2,\,\ld\,, (2l-k)\,$.  It should also be noted that, unlike in the 
case of the $su(2)$ algebra, the quadratic algebra (\ref{cqa}) has infinitely 
many inequivalent unitary irreducible representations of the same dimension. 
For example, for each value of $k = 1/2,1,3/2,\ld\,,$ there is a   
$2$-dimensional representation of the algebra (\ref{cqa}) given by 
\bea 
Q_0 & = & \fr{1}{2} \lrb \ba{cc}
                     k-1 & 0   \\
                     0   & k+1 \ea \rrb\,, \quad 
Q_+ = \lrb \ba{cc} 
                   0 & 0 \\
           \sqrt{2k} & 0 \ea \rrb\,, \quad 
Q_- = \lrb \ba{cc}
          0 & \sqrt{2k}  \\
          0 & 0          \ea \rrb\,, \nn \\ 
\ck & = & k(1-k)\,, \quad 
\cl = l = \frac{1}{2}(k+1)\,, \quad   
\cc = \frac{1}{8}\lrb -3k^3 - 5k^2 + 11k - 3 \rrb\,, \nn \\
  &   &  
\eea
as can be verified directly. \\   

As before, let us make the association 
\be
\lmd k , l , n \rra \lra \fr{z_2^{2k-1} z_3^{2l-k} \lrb z_1z_2/z_3 \rrb^n}
                            {\sqrt{n!(n+2k-1)!(2l-k-n)!}}\,. 
\ee
Since $k$ and $l$ are constants for a given representation we can take 
\be
\phi_{k,l,n}(z) = \fr{z^n}{\sqrt{n!(n+2k-1)!(2l-k-n)!}}\,, \quad 
  n = 0,1,2,\,\ld\,,(2l-k)\,, 
\ee
as the set of basis functions for the single variable realization 
\be 
Q_0 = z \ddz + k - l\,, \quad 
Q_+ = -z^2 \ddz + (2l-k)z\,, \quad 
Q_- = z \ddzt + 2k \ddz\,, 
\ee
leading to the representation (\ref{mrkl}).  

\section{Three-mode construction of noncompact quadratic algebras} 
\renewcommand{\theequation}{\arabic{section}.{\arabic{equation}}}
\setcounter{equation}{0}

\noindent
Let us choose 
\bea
Q_0 & = & \fr{1}{4} \lrb a_1^\da a_1 + a_2^\da a_2 + 2a_3^\da a_3 + 1 \rrb 
      = \fr{1}{2} \lrb K_0 + a_3^\da a_3 \rrb\,, \nn \\ 
Q_+ & = & a_1^\da a_2^\da a_3^\da = K_+ a_3^\da\,, \quad 
  Q_- = Q_+^\da = a_1 a_2 a_3 = K_- a_3\,, \nn \\
\ck & = & \fr{1}{4} \lsb 1 - \lrb a_1^\da a_1 - a_2^\da a_2 \rrb^2 \rsb 
      = K^2\,, \nn \\
\cl & = & \fr{1}{4} \lrb a_1^\da a_1 + a_2^\da a_2 - 2a_3^\da a_3 + 1 \rrb 
      = \fr{1}{2} \lrb K_0 - a_3^\da a_3 \rrb\,, 
\eea
where $\lrb K_0, K_+, K_- \rrb$ generate $su(1,1)$ with $K^2$ as the Casimir 
operator.  Now, the algebra becomes 
\bea 
\lsb \ck , \cl \rsb & = & 0\,, \quad
\lsb \ck , Q_{0,\pm} \rsb = 0\,, \quad
\lsb \cl , Q_{0,\pm} \rsb = 0\,, \nn \\ 
\lsb Q_0 , Q_\pm \rsb & = & \pm Q_\pm\,, \quad 
\lsb Q_+ , Q_- \rsb = -3Q_0^2 - (2\cl + 1)Q_0 - (\ck - \cl(\cl - 1))\,. \nn \\ 
   &   &  
\lb{ncqa}
\eea 
The Casimir operator of this algebra is given by 
\be
C = Q_+Q_- - Q_0^3 + (\cl - 1)Q_0^2 - \lrb \ck - \cl^2 \rrb Q_0 
\ee 
apart from additional constant functions of $\ck$ and $\cl$.  \\ 

The constancy of $\ck$ and $\cl$ in an irreducible representation leads to the 
choice of basis consisting of the infinite three-mode Fock states 
\be
\lmd k , l , n \rra = \lmd n, n+2k-1, n+k-2l \rra\,, \quad 
 n = 0, 1, 2,\,\ld\,, 
\lb{ncqabs}
\ee
with 
\be
k-2l = 0,1,2,\ld\,, \quad 
k = 1/2, 1, 3/2, \ld\,, 
\ee
and 
\be
\ck \lmd k,l,n \rra = k(1-k) \lmd k,l,n \rra\,, \qquad
\cl \lmd k,l,n \rra = l \lmd k,l,n \rra\,. 
\ee 
The set of basis states (\ref{ncqabs}) carry the infinite dimensional unitary 
irreducible representation of the quadratic algebra (\ref{ncqa}) given by 
\bea 
Q_0 \lmd k,l,n \rra & = & (k-l+n) \lmd k,l,n \rra\,, \nn \\ 
Q_+ \lmd k,l,n \rra & = & \sqrt{(n+1)(n+2k)(n+k-2l+1)}\,\lmd k,l,n+1 \rra\, 
\nn \\  
Q_- \lmd k,l,n \rra & = & \sqrt{n(n+2k-1)(n+k-2l)}\,\lmd k,l,n-1 \rra\,, \nn \\ 
\ck \lmd k,l,n \rra & = & k(1-k) \lmd k,l,n \rra\,, \qquad  
\cl \lmd k,l,n \rra = l \lmd k,l,n \rra\,, \nn \\
   &   & \qquad n = 0,1,2,\ld\,. 
\lb{ncqarep}
\eea 
The Casimir operator has the value $l\lrb l-k^2 \rrb$ in this $(k,l)$-th 
representation. As in the compact case, to obtain this representation one can 
also use an alternative set of basis states 
$\lmd k , l , n \rra = \lmd n, n+2k-1, n+k-2l \rra$\,, with  
$n = 0, 1, 2,\,\ld$\,. \\ 

From the association 
\be 
\lmd k , l , n \rra \lra \fr{\lrb z_1 z_2 z_3 \rrb^n z_2^{2k-1} z_3^{k-2l}} 
                         {\sqrt{n!(n+2k-1)!(n+k-2l)!}} 
\ee 
it is clear that we can take 
\be 
\psi_{k,l,n}(z) = \fr{z^n}{\sqrt{n!(n+2k-1)!(n+k-2l)!}} 
\ee
as the set of basis functions for a single variable realization of the algebra 
(\ref{ncqa}).  The corresponding realization is 
\bea
Q_0 & = & z \ddz + k - l\,, \qquad Q_+ = z\,, \nn \\  
Q_- & = & z^2 \ddzth + (3k-2l+2) z \ddzt + \lrb 2k^2-4kl+2k \rrb \ddz\,. 
\eea 

\section{Coherent states of the quadratic algebras} 
\renewcommand{\theequation}{\arabic{section}.{\arabic{equation}}}
\setcounter{equation}{0}

\noindent
As is well known, coherent states form an overcomplete set of nonorthogonal  
states, generally labeled by a continuous index, say $\al$, providing a 
resolution of identity $\int \lmd \al \rra \lla \al \rra d\al = I$ and hence 
a useful set of basis states.  Coherent states associated with the unitary 
irreducible representations of the dynamical algebra of a physical system are 
very useful in certain studies of the system.  \\

For the noncompact algebra (\ref{ncqa}) the Barut-Girardello type coherent 
states associated with the $(k,l)$-th irreducible representation can be 
defined as the eigenstates of $Q_-$\,:
\be
Q_- \lmd k,l,\al \rra = \al \lmd k,l,\al \rra\,, \qquad
\lmd k,l,\al \rra = \sum_{n=0}^\infty\,c_n(\al) \lmd k,l,n \rra\,,
\lb{cse}
\ee
where the complex $\al$ labels these states.  Solving for $\lmd k,l,\al \rra$ 
using the representation (\ref{ncqarep}) we get easily 
\bea
\lmd k,l,\al \rra & = & 
        \lsb \fr{(2k)!(k-2l+1)!}{{}_0F_2(-;2k,k-2l+1;|\al|^2)} \rsb^{\fr{1}{2}}
        \nn \\
  &   & \times \sum_{n=0}^\infty\,\fr{\al^n}
                       {\sqrt{n!(n+2k-1)!(n+k-2l)!}}\,\lmd k,l,n \rra\,.
\eea 

In terms of the single variable realization the coherent state equation  
(\ref{cse}) can be written as 
\bea 
  &   &  \lsb z^2 \ddzth + (3k-2l+2) z \ddzt 
         + \lrb 2k^2-4kl+2k \rrb \ddz \rsb \Psi_{k,l}(\al, z) \nn \\ 
  &   &  \qquad \qquad \qquad \qquad \qquad \qquad 
         = \al \Psi_{k,l}(\al, z)\,,
\eea
which is the equation for  
\be
\Psi_{k,l}(\al, z) = {}_0F_2(-;2k,k-2l+1;\al z)\,. 
\ee
The resolution of the identity is given by:
\begin{equation}
\int d \sigma(\alpha,\alpha^{*};k,l) |\alpha;k,l \rangle\langle \alpha;k,l| = \hat{1} .
\end{equation}

With a polar decomposition ansatz $\alpha=re^{i\theta}$ we get,
 \be
d\sigma(\alpha^{*},\alpha;k,l) =N_{l,k}(r^2)( \sigma(r^2)) d\theta r dr ,
\ee
where
\be
N_{l,k}(r^2)=_0F_2(-;2k,k-2l+1;r^2)
\ee
The integral (5.5) reduces to the following condition on $\sigma(r^2)$:
\begin{equation}
\frac{1}{2}\int_{0}^{\infty} d(r^2)\, \sigma(r^2) \, (r^2)^{(n+1)-1} = \frac{1}{2\pi}\Gamma(n+1)
\, \frac{\Gamma(2k+n)
\Gamma(k-2l+1+n)}
{\Gamma(2k)\Gamma(k-2l+1)}.
\end{equation}
 $\sigma(r^2)$ is found by an inverse Mellin transform to be:
\be
\sigma(r^2)=\fr{1}{ \pi \Gamma(2k)(\Gamma(k-2l+1 )}
G^{3\,0}_{0\,3} (r^2|^{-}_{0,k-2l+1,2k} ),
\ee
where $G$ is the Meijers G function \cite{dbm}.

Another useful set of coherent states for the quadratic algebras will be the 
analogues of the well known Perelemov-type coherent states.  A technique for 
generalizing the Perelemov-type construction of coherent states for the 
nonlinear algebras, using a mapping of the given compact or noncompact algebra 
to $su(2)$ or $su(1,1)$ respectively, has been presented in detail in our 
earlier work \cite{Su}. For completeness we give the expressions for these states.
For the non compact case the {\it Perelomov type} states are:
\be
|\beta=N\sum_{n} (\beta)^n \sqrt{\frac{\Gamma(2k+n) \Gamma(k-2l+1+n)}{\Gamma(n+1)\Gamma(2k)\Gamma(k-2l+1)}}|l,k.n>,
\ee
where the normalization coefficient N is given by
\be
N=(^2F_0(2k,k-2l+1;(|\beta|^2)))^{-\frac{1}{2}}.
\ee
The resolution of the identity in this case reduces to finding $\sigma(r^2)$ such that,
\begin{equation}
\int_{0}^{\infty} d(r^2)\, \sigma(r^2) \, (r^2)^{(n+1)-1} = \frac{1}{\pi}\Gamma(n+1)
\, \frac{\Gamma(2k)\Gamma(k-2l+1)}
{\Gamma(2k+n)\Gamma(k-2l+1+n)}
\end{equation}
and the resultant $\sigma(r^2)$ is given by
\be
\sigma(r^2)=\fr{1}{ \pi} (\Gamma(k-2l+1 )\Gamma(2k)
G^{1,0}_{2,1} (r^2|^{k-2l+1,2k-1}_{0} ),
\ee
where $G$ is the Meijers G function.

The corresponding states for the compact case are given by:
\be
|\alpha,k,l>=N\sum_{n=0}^{2l-k} (\alpha)^n \sqrt{\frac{(2l-k)! (2k-1+n)!}{n!(2l-k-n)!(2k-1)!}}|l,k.n>.
\ee
For the purposes of calculating the measure for the resolution of identity we define
$\gamma=\frac{1}{\alpha}$.
The coherent state $|\gamma,k,l>$ becomes
\be
|\gamma,k,l>=N \gamma^{k-2l}\sum_{n=0}^{2l-k} (\gamma)^n \sqrt{\frac{(2l-k)! (k+2l-1-n)!}{n!(2l-k-n)!(2k-1)!}}|l,k.n>,
\ee
with the normalization coefficient N  given by:
\be
N=( \frac{\Gamma(k+2l)}{(|\gamma|^2)^{2l-k}}\Phi(k-2l,1-2l-k;(|\gamma|^2)))^{-\frac{1}{2}},
\ee
where $\Phi(a,b,x)$ is the Confluent Hypergeometric function $ (^1F_1)$.

The resolution of the identity
for the coherent states $|\gamma ,k,l\rangle$ is given by
\begin{equation}
\int d \mu(\gamma,\gamma^{*};k,l) |\gamma;k,l \rangle\langle \gamma;k,l| = \hat{1}_{l,k} , \label{id}
\end{equation}
where $\hat{1}_{l,k}$ is the projection operator on the subspace 
${\cal H}_{2l-k}$:
\begin{equation}
\hat{1}_{l,k} = \sum_{n=0}^{2l-k} |l,k,n \rangle \langle l,k,n| .
\end{equation}

Again defining $\gamma=re^{i\theta}$
we have,
\begin{equation}
\sum_{n=0}^{2l-k} \frac{ \Gamma(2l+k-n) }{n! (2l-k-n)! } \left[ 
\int_{0}^{\infty}( r^2)^{n} M(r^2;k,l) d (r^2) \right] 
|l,k,n \rangle \langle l,k,n |  
= \hat{1}_{l,k} , \nonumber
\end{equation}
where we have defined
\begin{equation}
M(r^2;k,l) \equiv \frac{\pi (2l-k)!}{\Gamma(2l+k)} \frac{ \sigma(r^2;k,l) }{
(\Phi(k-2l,1-2l-k;r^2))}
  .
\end{equation}
By using the integral \cite{Erd}, 
\begin{equation}
\int_{0}^{\infty} r^{b-1} \Phi(a;c;-r) d r = \frac{ \Gamma(b)
\Gamma(c) \Gamma(a-b) }{ \Gamma(a) \Gamma(c-b) } ,
\end{equation}
we obtain 
\[
M(r^2;k,l) = \frac{ \Gamma(2l-k+2) }{ \Gamma(2l+k+1) }\, 
\Phi(2l-k+2;2l+k+1;-r^2) .
\]

This gives us the final expression for the integration measure: 
\begin{equation}
 d \mu(\gamma,\gamma^*;k,l) = \frac{1}{2\pi} \frac{(2l-k+1)}{(2l+k+1) } 
\Phi(k-2l;1-2l-k;r^2) 
\Phi(2l-k+2;k+2l+1;-r^2) d (r^2) d \theta .
\end{equation}

The resolution of the identity is important because it allows the use
of the coherent states as a basis in the state space. All three sets of coherent states
can also be easily shown to be overcomplete \cite{Su}.

\section{Concluding remarks\,: Physical applications} 
\renewcommand{\theequation}{\arabic{section}.{\arabic{equation}}}
\setcounter{equation}{0}

\noindent
As already mentioned in the introduction the nonlinear algebras are finding 
several applications in physical problems. Here we shall like to make some 
observations with reference to the bosonic constructions of the quadratic 
algebras we have presented.  \\

Let us consider the Hamiltonian 
\be
H = a_1^\da a_1 + a_2^\da a_2 + 2a_3^\da a_3 
\lb{anio}
\ee 
which describes, with units $\hbar = 1$ and $\omega = 1$, and apart from the 
additional zero-point level constant, a three-dimensional anisotropic quantum 
harmonic oscillator with the frequency in the third direction twice that in 
the perpendicular plane.  From the construction of the compact quadratic 
algebra given above we recognize that 
\be
H = 4\cl - 1\,, 
\ee
where $\lrb \cl , \ck , Q_0, Q_\pm \rrb$ generate the quadratic algebra 
(\ref{cqa}).  Thus, $\lrb \ck , Q_0 , Q_\pm \rrb$ are constants of motion 
for the system (\ref{anio}) and the quadratic algebra (\ref{cqa}) is its 
dynamical algebra.  Since $\cl$ has the spectrum 
\be
\cl = l = n/4\,, \qquad n = 1,2,3,\ld\,, 
\ee
it is clear that the Hamiltonian (\ref{anio}) has the spectrum 
\be
H = N\,, \qquad N = 0,1,2,\ld\,. 
\ee
Each level can be labeled by the eigenvalues of the complete set of commuting 
operators $\lrb H , \ck, Q_0 \rrb$.  \\

It is interesting to compute the degeneracy of the $N$-th level using the 
representation theory of the algebra (\ref{cqa}).  For the $N$-th level the 
value of $\cl$ is $l = (N+1)/4$.  Calculating the corresponding values of 
$k$ for which finite dimensional representations are possible we find that 
the dimensions of the associated irreducible representations are 
$(1,2,\ld\,,2m+1)$ if $N = 4m$ or $4m+1$\,, and $(1,2,\ld\,,2m+2)$ if 
$N = 4m+2$ or $4m+3$.  The degeneracy of the level is the sum of the 
dimension of the $k = 1/2$ representation and twice the dimensions of 
$k > 1/2$ representations.  One has to count the dimensions of $k > 1/2$ 
representations twice in the sum since there are two possible choices for 
the bases leading to the same representation in these cases as already 
noted.  Now, the four cases, $N = 4m, 4m+1, 4m+2, 4m+3$, are to be considered 
separately.  The result is\,: the degeneracies of the levels, 
$N = 4m, 4m+1, 4m+2, 4m+3$, resectively, are $(2m+1)^2$, $(2m+1)(2m+2)$, 
$4(m+1)^2$ and $2(m+1)(2m+3)$\,.  In other words, the number of compositions of 
the integer $N$ (partitions with ordering taken into account) in the prescribed 
pattern $n_1 + n_2 + 2n_3$, with the interchange of $n_1$ and $n_2$ taken into 
account, is $(2m+1)^2$, $(2m+1)(2m+2)$, $4(m+1)^2$ and $2(m+1)(2m+3)$, if 
$N = 4m, 4m+1, 4m+2$, and $4m+3$, resectively.  \\

It is to be noted that in the above example the sum of all the dimensions of the 
irreducible representations associated with the given $l = (N+1)/4$ gives the 
number of partitions of $N$ in the pattern $n_1 + n_2 + 2n_3$, disregarding the 
interchange of $n_1$ and $n_2$.  This leads to the result that the number of 
such partitions is $(m+1)(2m+1)$ for $N = 4m$ or $4m+1$ and $(m+1)(2m+3)$ for 
$N = 4m+2$ or $4m+3$.  Thus, it is interesting to observe this connection 
between the quadratic algebra and the theory of partitions.  \\

An another interesting possibility is suggested by the structure of the compact 
quadratic algebra (\ref{cqa}).  Let us define 
\be
N = Q_0\,, \quad
A = \frac{1}{\sqrt{\cl(\cl+1)-\ck}}\,Q_-\,, \quad
A^\da = \frac{1}{\sqrt{\cl(\cl+1)-\ck}}\,Q_+\,.
\ee
The the algebra (\ref{cqa}) becomes 
\bea
\lsb N , A \rsb & = & -A\,, \quad 
\lsb N , A^\da \rsb = A^\da\,, \nn \\ 
\lsb A , A^\da \rsb & = & 1 - \fr{3}{\cl(\cl+1)-\ck}\,N^2         
                        - \fr{2\cl-1}{\cl(\cl+1)-\ck}\,N\,. 
\lb{qo}
\eea 
We may consider this as the defining algebra of a quadratic oscillator, 
corresponding to a special case of the general class of deformed oscillators 
(\cite{Ar}-\cite{D})\,:
\be
\lsb N , A \rsb = -A\,, \quad 
\lsb N , A^\da \rsb = A^\da\,, \quad
\lsb A , A^\da \rsb = F(N)\,.
\ee
The quadratic oscillator (\ref{qo}) belongs to the class of generalized 
deformed parafermions \cite{Q}.  It should be interesting to study the 
physics of assemblies of quadratic oscillators.  In fact, the canonical 
fermion, with 
\be 
N = \lrb \ba{cc}
         0 & 0  \\
         0 & 1 \ea \rrb\,, \quad 
f = \lrb \ba{cc} 
         0 & 1 \\
         0 & 0 \ea \rrb\,, \quad 
f^\da = \lrb \ba{cc}
          0 & 0  \\
          1 & 0  \ea \rrb\,, 
\ee
is a quadratic oscillator!  Observe that 
\be
\lsb N , f \rsb = -f\,, \quad 
\lsb N , f^\da \rsb = f^\da\,, \quad
\lsb f , f^\da \rsb = 1 - \fr{1}{2} N - \fr{3}{2} N^2\,.
\ee

To conclude, we find that the study of polynomial algebras should lead to 
interesting connections with the theory of partitions and new physical systems 
in which the polynomial oscillators replace the role of the canonical 
oscillators, besides having the other known applications.

\end{document}